\documentclass[aps,prb,twocolumn,showpacs]{revtex4}
\usepackage{graphicx}

\begin{document}

\title{Photoemission and x-ray absorption spectroscopy study of
electron-doped colossal magnetoresistance manganite:
La$_{0.7}$Ce$_{0.3}$MnO$_3$ film}

\author{S. W. Han$^1$, J.-S. Kang$^{2,*}$, K. H. Kim$^1$, 
	J. D. Lee$^1$, J. H. Kim$^2$, S. C. Wi$^2$, 
	C. Mitra$^3$, P. Raychaudhuri$^4$, S. Wirth$^5$, 
	K. J. Kim$^6$, B. S. Kim$^6$, J. I. Jeong$^7$,
	S. K. Kwon$^8$, and B. I. Min$^8$}

\affiliation{$^1$Department of Physics and the Research Institute of
	Natural Sciences, Gyeongsang National University, 
	Chinju 660-701, Korea}

\affiliation{$^2$Department of Physics,
	The Catholic University of Korea, Puchon 420-743, Korea}

\affiliation{$^3$Department of Materials Science, University of 
	Cambridge, Pembroke St., Cambridge CB2 3QZ, UK}

\affiliation{$^4$Tata Institute of Fundamental Research, Homi Bhabha 
	Road, Bombay 400005, India} 

\affiliation{$^5$Max Planck Institute for Chemical Physics of Solids, 
	N\"{o}thnizer Strasse 40, 01187 Dresden, Germany}

\affiliation{$^6$Pohang Accelerator Laboratory (PAL), 
	Pohang University of Science and Technology, 
	Pohang 790-784, Korea}

\affiliation{$^7$Research Institute of Industrial Science and Technology, 
	Pohang 790-600, Korea}

\affiliation{$^8$Department of Physics, 
	Pohang University of Science and Technology, 
	Pohang 790-784, Korea}

\date{\today}

\begin{abstract}

The electronic structure of La$_{0.7}$Ce$_{0.3}$MnO$_{3}$ (LCeMO) thin 
film has been investigated using photoemission spectroscopy (PES) 
and x-ray absorption spectroscopy (XAS). The Ce $3d$ core-level PES 
and XAS spectra of LCeMO are very similar to those of CeO$_2$,
indicating that Ce ions are far from being trivalent.
A very weak $4f$ resonance is observed around the Ce $4d \rightarrow 
4f$ absorption edge, suggesting that the localized Ce $4f$ states are 
almost empty in the ground state. The Mn $2p$ XAS spectrum reveals 
the existence of the Mn$^{2+}$ multiplet feature, confirming 
the Mn$^{2+}$-Mn$^{3+}$ mixed-valent states of Mn ions in LCeMO.
The measured Mn $3d$ PES/XAS spectra for LCeMO agrees reasonably well 
with the calculated Mn $3d$ PDOS using the LSDA+$U$ method.
The LSDA$+U$ calculation predicts a half-metallic ground state for LCeMO.

\end{abstract}
 \maketitle

\section{Introduction}

Perovskite Mn oxides of R$_{1-x}$A$_x$MnO$_3$ (RAMO; R:rare earth;
A:divalent cation) \cite{Jin94} have attracted much attention
due to the colossal-magnetoresistance (CMR) behavior. 
Pure LaMnO$_3$ is an antiferromagnetic insulator.
When LaMnO$_3$ is doped with divalent cations, it undergoes
a phase transition to a ferromagnetic metal, in which Mn ions can 
exist in the formally trivalent and tetravalent states. 
Zener has explained \cite{Zener51} the simultaneous metallic and 
ferromagnetic transition in RAMO in terms of the double-exchange (DE) 
interaction between spin-aligned Mn$^{3+}$ ($t_{2g}^{3}e_{g}^{1}$) 
and Mn$^{4+}$ ($t_{2g}^{3}$) ions through oxygen ions.

In the DE model, Mn ions should exist in mixed-valent states 
to maintain the correlation between magnetism and conductivity.
The question has been raised whether the DE mechanism is still 
operative when a tetravalent ion  
is doped instead of a divalent ion \cite{Mandal97}.
This will result in a system with  
Mn ions being in the Mn$^{2+}$ ($t_{2g}^3e_{g}^2$)/Mn$^{3+}$ 
($t_{2g}^3e_{g}^1$) mixed-valent states.
Interestingly, the metal-insulator (M-I) and ferromagnetic transitions
and the concomitant CMR phenomenon have been observed in the Ce-doped 
manganites of R$_{0.7}$Ce$_{0.3}$MnO$_3$ (R=La, Pr, Nd) 
\cite{Mandal97,Gebh99}.
If Ce ions in R$_{0.7}$Ce$_{0.3}$MnO$_3$ exist in the tetravalent 
states, Mn$^{2+}$ ions could be formed and electron-like charge 
carriers would be responsible for the metallic conductivity and 
ferromagnetism. Therefore it is essential to know the electronic 
structures of R$_{0.7}$Ce$_{0.3}$MnO$_3$ in order to understand 
the underlying physics for the metallic ferromagnetism properly.

In our previous photoemission spectroscopy (PES) study on 
polycrystalline La$_{0.7}$Ce$_{0.3}$MnO$_{3}$ (LCeMO) bulk samples
\cite{Kang01}, we have found that Ce ions in LCeMO are mainly 
in the tetravalent ($4+$) states, which allows the existence 
of the divalent Mn$^{2+}$ ions in LCeMO. 
However, one of the main difficulties in this system is that 
the La$_{1-x}$Ce$_x$MnO$_3$ system forms in the single phase
only in epitaxial thin films \cite{Ray99,Mitra01}.
Recently, the Ce$^{4+}$ valence state and the Mn$^{2+}$-Mn$^{3+}$ 
mixed-valent states were observed in a thin film of LCeMO
\cite{Mitra03} via x-ray absorption spectroscopy (XAS).
Nevertheless, the detailed spectroscopic information on the electronic
states near the Fermi energy $\rm E_F$ is lacking for LCeMO,
which is important in understanding the nature of the charge carriers.
In this paper, we report the PES  and XAS study of LCeMO thin films. 
This work includes the resonant photoemission spectroscopy (RPES) 
measurement near the Ce $4d \rightarrow 4f$ absorption edge, and
the XAS measurements near the Ce and La $3d$, Mn $2p$, and O $1s$ 
absorption edges.
PES and XAS data have been compared to the band-structure calculations
performed in the LSDA$+U$ method (LSDA: local spin-density 
approximation) where $U$ denotes the on-site Coulomb correlation
interaction for both Mn $3d$ and Ce $4f$ electrons. 

\section{Experimental and calculational details}

The epitaxial films of La$_{0.7}$Ce$_{0.3}$MnO$_3$ were deposited 
on LaAlO$_3$ (LAO) substrates using a KrF excimer laser, as 
described in Ref.~\cite{Mitra01}. 
PES and XAS measurements were performed at the 2B1 beamline 
of the Pohang Accelerator Laboratory (PAL). 
The chamber pressure was about $\sim 5 \times$ 10$^{-10}$ Torr 
during measurements. 
All spectra were obtained at room temperature. 
The Fermi level of the system was determined from the valence-band 
spectrum of a Au foil in electrical contact with samples. 
The overall instrumental resolution was about $\sim 200$ meV 
at a photon energy $h\nu\approx 30$ eV and $\sim 300$ meV at
$h\nu \approx 120$ eV. 
To obtain clean surfaces, the samples were annealed repeatedly at 
$\sim$ 700 $^o$C in the O$_2$ pressure of about $1 \times 10^{-8}$ Torr. 
The cleanliness of sample surfaces was monitored by the absence of
the bump around 9 eV binding energy (BE) and the symmetrical line 
shape of the O $1s$ core level. 

The electronic structures of LCeMO have been calculated by employing 
the self-consistent LMTO (linearized muffin-tin-orbital) band method.
To simulate the Ce-doped LaMnO$_3$ system, 
we considered a supercell of La$_2$CeMn$_3$O$_9$
with a tetragonal structure ($a=3.873 \AA$ and $c=11.618 \AA$).
The partial densities of states (PDOSs) for LCeMO 
were obtained from the LSDA+$U$ band method which incorporates 
the spin-orbit (SO) interaction \cite{Kwon00}.

\section{Result and discussion}

We first present the Ce and La $3d$ core level PES and XAS spectra
of LCeMO. XAS and core-level spectroscopy are powerful methods 
of determining
the valence states of ions in solids. In the final state of core-level
PES and XAS spectra, a core hole is left behind, and it couples 
with valence electrons.
In the systems with incompletely filled $4f$ or $3d$ electrons, 
the coupling between the core hole and $4f$ or $3d$ electrons is 
strong enough to cause the characteristic spectral splitting in XAS 
and XPS spectra. Therefore, by analyzing XPS and XAS, 
an important information on 
the $4f$ and $3d$ valence electrons can be obtained. 

\begin{figure}[t]
\includegraphics[scale = 0.6]{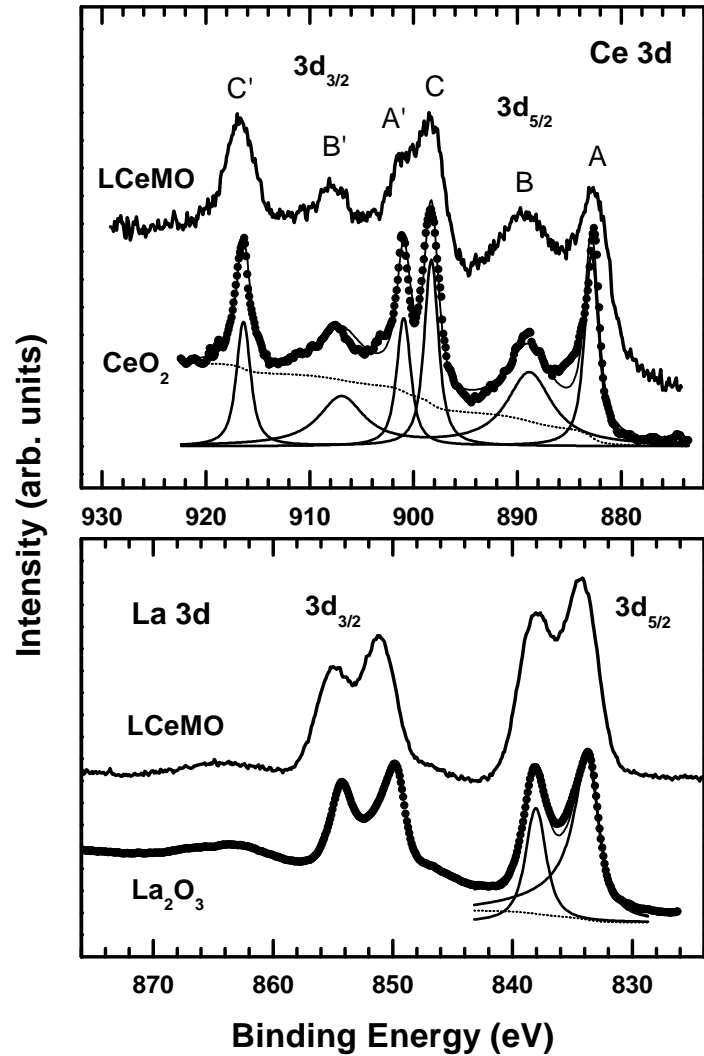}
\caption{Top: Comparison of the Ce $3d$ core-level PES spectra of LCeMO 
	and CeO$_{2}$ (from Ref.~\protect\cite{Wuill84}).
	The curve fitting results (solid lines) of 
	the Ce $3d$ spectrum for CeO$_2$ are superposed 
	on the measured spectrum (dots). 
	Bottom: Similarly for the La $3d$ core-level PES spectra 
	for LCeMO and La$_2$O$_3$ (from Ref.~\protect\cite{Suzuki00}).}
\label{3d}
\end{figure}

Figure~\ref{3d} compares the Ce $3d$ core-level PES spectrum of LCeMO 
to that of CeO$_2$ with formally tetravalent Ce ions (Ce$^{4+}$) 
(upper panel), and the La $3d$ core-level PES spectrum of LCeMO to that
of La$_2$O$_3$ with formally trivalent La ions (La$^{3+}$) (bottom 
panel).
The $3d$ core-level spectra of CeO$_{2}$ and La$_2$O$_3$ were 
reproduced from Ref.~\cite{Wuill84} and Ref.~\cite{Suzuki00},
respectively. 
The spectrum of CeO$_2$ was then shifted so that the $3d_{5/2}$ 
main peak is aligned to that of LCeMO. 
These spectra exhibit the same magnitudes of the spin-orbit splittings
between the $3d_{3/2}$ and $3d_{5/2}$ levels, 16.8 eV for the La $3d$ 
spectra and 18.1 eV for the Ce $3d$ spectra, respectively. 
It is clearly observed that the Ce and La $3d$ spectra of LCeMO are 
very similar to those of CeO$_{2}$ and La$_2$O$_3$, respectively,
indicating that Ce ions in LCeMO are nearly tetravalent (Ce$^{4+}$)
while La ions are trivalent (La$^{3+}$). 

The solid lines along the measured $3d$ spectra for CeO$_2$ and 
La$_2$O$_3$ denote the curve-fitting results, by employing 
the Doniach-Sunjic line-shape function \cite{DS76}. 
The curve-fitting analysis reveals that the Ce $3d$ PES spectrum 
of CeO$_2$ consists of six peaks, and similarly for LCeMO. 
These six peaks arise from the three-peak structures for each 
spin-orbit split component of $3d_{5/2}$ and $3d_{3/2}$, respectively. 
Among the three-peak structures, the highest BE components 
(C, C$^{\prime}$) correspond to the roughly $3d^9 4f^0$ final-state 
configuration \cite{Wuill84,Fuji83}. 
So this figure indicates that CeO$_2$ and LCeMO have a very large
amount of $3d^{10} 4f^0$ initial-state configuration.
According to the impurity Anderson Hamiltonian (IAH) analysis 
\cite{Wuill84,Kotani92}, 
the lowest BE peaks (A, A$^{\prime}$) and the middle BE peaks 
(B, B$^{\prime}$) correspond to the bonding and anti-bonding states
of the strongly mixed $3d^9 4f^1 \b{L}$ and $3d^9 4f^2 \b{L}^2$ 
final-state configurations ($\b{L}$: a ligand hole).
Since the charge transfer energy $\Delta_f$ between the $4f^0$ 
and $4f^1 \b{L}$ configurations is small in CeO$_2$,
the ground state is strongly mixed between $4f^0$ and $4f^1 \b{L}$ 
configurations. 
Hence the valence electronic states in CeO$_2$ contain non-negligible 
contributions from the extended states of $f$ symmetry, even though 
the localized $4f$ states remain nearly unoccupied. This results 
in the average $4f$ electron number $n_f$ of about $0.5$ in CeO$_2$.

The La $3d$ spectra show the double-peak structures of nearly 
equal intensity for both the $3d_{5/2}$ and $3d_{3/2}$ levels. 
Based on the IAH analysis \cite{Wuill84,Kotani92}, it is well known 
that the La $4f$ level in the ground state of La$_2$O$_3$ 
is nearly empty due to the very weak hybridization between La $4f$ 
and O $2p$ orbitals. On the other hand,
in the $3d$ core-hole final state, the attractive Coulomb interaction 
$U_{fc}$ between the $3d$ core hole and the $4f$ electrons ($U_{fc}<0$) 
pulls down the $4f$ level, so that the charge transfer energy 
$\Delta_f$ between $3d^9 4f^0$ and $3d^9 4f^1 \b{L}^1$ becomes almost 
vanishing, resulting in a strong hybridization  between $3d^9 4f^0$ 
and $3d^9 4f^1 \b{L}^1$ final-state configurations.
Therefore the $3d^9 4f^0$ and $3d^9 4f^1 \b{L}^1$ configurations are 
strongly mixed in the $3d$ core-hole final state of La$_2$O$_3$.
The two peaks of the La $3d$ core level PES spectrum in La$_2$O$_3$ 
correspond to the bonding and antibonding states of the 
$3d^9 4f^0$ and $3d^9 4f^1 \b{L}^1$ configurations.
Due to the strong final-state mixing in La$_2$O$_3$, 
the intensities \cite{int} of the two peaks become comparable.
This interpretation can be similarly applied to the La $3d$ PES 
spectrum of LCeMO.  Thus, Fig.~\ref{3d} provides  
evidence that La $4f$ states are almost unoccupied in the ground state 
($4f^0$), but are strongly hybridized with the O $2p$ states
in the $3d$ core-hole final state.

\begin{figure}[t]
\includegraphics[scale = 0.6]{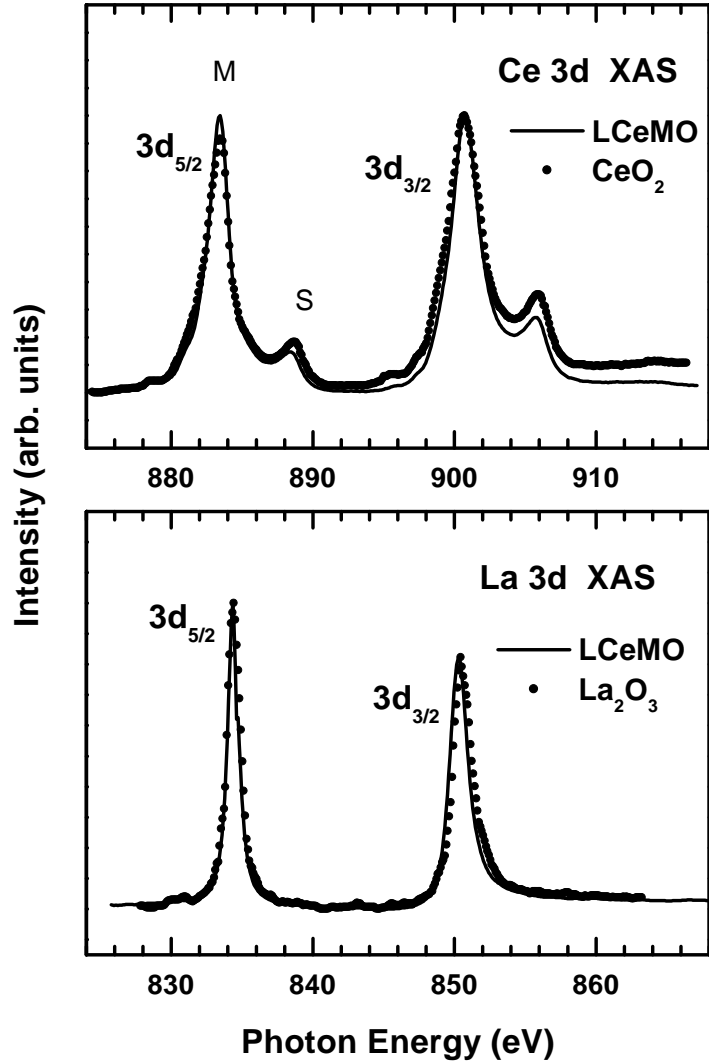}
\caption{Top: Comparison of the Ce $3d$ XAS spectra of LCeMO 
	and CeO$_2$ (from Ref.~\protect\cite{Kaindl84}).
	Bottom: Comparison of the La $3d$ XAS spectra of LCeMO
	and La$_2$O$_3$ (from Ref.~\protect\cite{Kaindl84}. }
\label{3dxas}
\end{figure}

Figure~\ref{3dxas} shows the Ce $3d$ XAS spectra of LCeMO and CeO$_2$, 
and the La $3d$ XAS spectra of LCeMO and La$_2$O$_3$. The $3d$ XAS 
spectra of both CeO$_2$ and La$_2$O$_3$ were reproduced from 
Ref.~\cite{Kaindl84}. 
The Ce $3d$ XAS spectrum of LCeMO is very similar to that of CeO$_{2}$,
consistent with the finding for Ce $3d$ PES spectrum, which confirms
the tetravalent valency of Ce ions in LCeMO. 
The Ce $3d$ XAS spectra of both LCeMO and CeO$_2$ reveal a main peak 
(M) and a weak satellite (S) on the high-energy side of the main peak. 
The main Ce $3d$ XAS peaks show no multiplet structures.
The main (M) and satellite (S) peaks correspond to the bonding and 
anti-bonding final states of the $3d^9 4f^1$ and $3d^9 4f^2 \b{L}^1$ 
mixed configurations. Since the ground state $|g \rangle$ of CeO$_2$ 
will be 
$|g \rangle \approx \alpha |f^0 \rangle + \beta |f^1 \b{L} \rangle$, 
the $3d$ XAS final states will have both the $3d^9 4f^1$ and 
$3d^9 4f^2 \b{L}$ configurations. In both LCeMO and CeO$_2$,
the energy difference between M and S in the Ce $3d$ XAS spectrum
($\approx 17.5$ eV) is approximately the same as that between 
A(A$^{\prime}$) and B(B$^{\prime}$) in the Ce $3d$ core-level PES 
spectrum ($\approx 18.1$ eV).
Note that, except for the weak satellite features, the Ce $3d$ XAS 
spectrum of CeO$_2$ is very similar to the La $3d$ XAS spectrum
of La$_2$O$_3$, suggesting that the multiplet structures of the $3d$
XAS of CeO$_{2}$ resemble those of La$_2$O$_3$ with La$^{3+}$ 
which has the $4f^0$ configuration in the ground state. 

\begin{figure}[t]
\includegraphics[scale = 0.53]{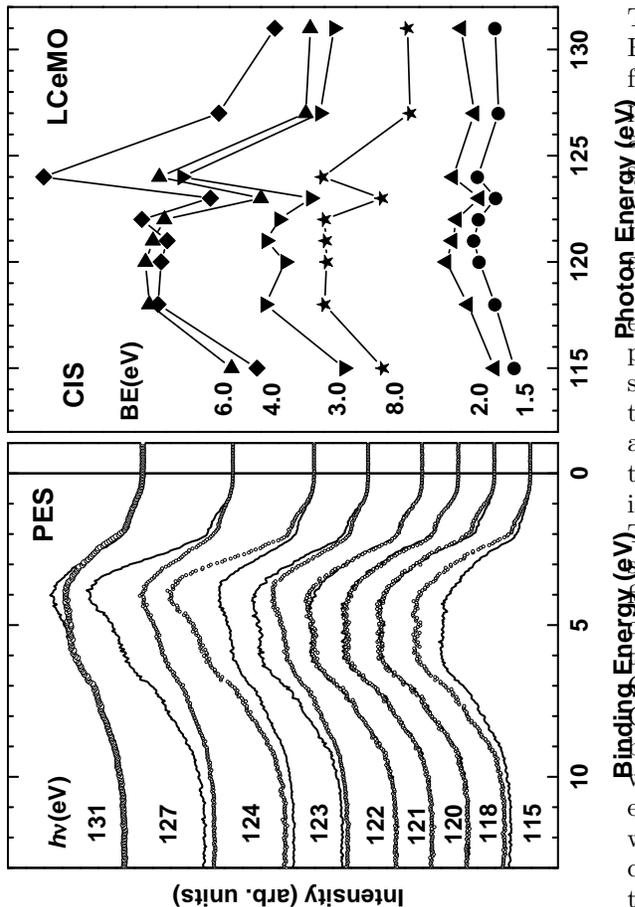}
\caption{Left: the valence-band spectra for LCeMO versus $h\nu$,
	obtained at the Ce $4d \rightarrow 4f$ absorption region. 
	In each set of the two superposed spectra, solid lines and open 
	circles correspond to the spectra for the lower $h\nu$ and 
	the next higher $h\nu$, respectively. 
	Each set of $h\nu$ values is labelled at the left 
	side of the spectra.
Right: the constant-initial-state (CIS) spectra of LCeMO for several
	initial states, which were obtained by plotting the 
	photoemission intensities at several BEs versus $h\nu$.  }
\label{rpes}
\end{figure}

Figure~\ref{rpes} presents the normalized valence-band spectra for
LCeMO obtained at the Ce $4d \rightarrow 4f$ absorption region.
In each set of the two superposed spectra, solid lines and open 
circles correspond to the spectrum for the lower $h\nu$ and that for 
the next higher $h\nu$. The values of $h\nu$'s are labelled at the left
side of the spectra in the increasing order from bottom toward top 
of the figure. If there are localized Ce $4f$ electrons, the $4f$ 
photoemission intensity is enhanced at $h\nu \approx 121$ eV due to 
the resonance effect through the interference between two processes.
The first is the direct photoemission process, such as 
\begin{equation}
4d^{10} 4f^n + h\nu \rightarrow 4d^{10} 4f^{n-1} \epsilon_k ~~~,
\nonumber
\end{equation}
where $\epsilon$$_k$ denotes the emitted electron. 
The second is the photoabsorption of a $4d$ electron 
to an unoccupied $4f$ state, followed by a two-electron super
Coster-Kronig decay, such as
\begin{equation}
4d^{10} 4f^n + h\nu \rightarrow 4d^{9} 4f^{n+1} \rightarrow
4d^{10} 4f^{n-1} \epsilon_k ~~~.
\nonumber
\end{equation}
The interference between these two processes leads to the Fano 
resonance.  
Such a RPES process will not be invoked for a tetravalent Ce$^{4+}$ 
ion ($4f^0$) because there is no direct process available.
Therefore the very weak enhancement around $h\nu\approx 121$ eV 
indicates that the localized Ce $4f$ states are nearly unoccupied 
in LCeMO and that the Ce valence is far from $3+$.
This finding is consistent with that for the Ce $3d$ core-level PES
and XAS spectra (Fig.~\ref{3d} and Fig.~\ref{3dxas}). 

\begin{figure}[t]
\includegraphics[scale = 0.55]{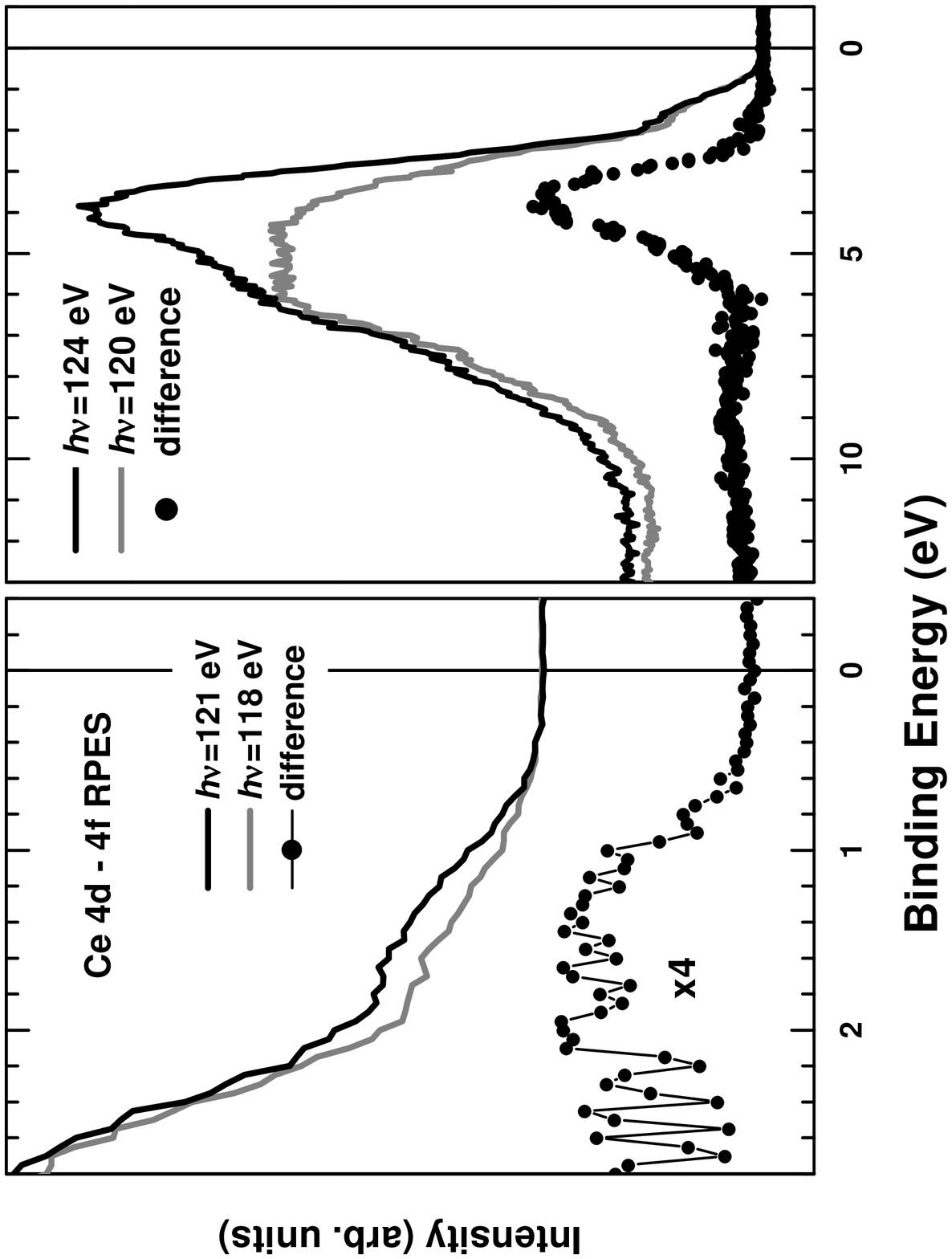}
\caption{Left: Comparison of the normalized valence-band PES spectra 
	of LCeMO obtained for $h\nu=118$ eV and $h\nu=121$ eV.
	The difference between these two spectra is shown at the bottom.
	Right: Similarly for $h\nu=120$ eV and $h\nu=124$ eV.  }
\label{diff}.
\end{figure}

The right panel of Fig.~\ref{rpes} shows $h\nu$ dependence 
of the emissions at several initial-state energies. 
These are the plots of the photoemission intensities at several BEs 
versus $h\nu$ around the Ce $4d$ absorption thresholds. Hence these 
curves measure the RPES cross-section line-shapes, and correspond to 
the constant-initial-state (CIS) spectra.
The vertical scale of this figure is arbitrary but it is the same 
for all the spectra with different BE states.
The O $2p$ states, with BE between $\sim 3 $ eV and $\sim 8$ eV, show 
the strong peaks around $\sim 124$ eV.
This peak reflects the resonance enhancement of the extended states 
of $f$ symmetry near the Ce $4d \rightarrow 4f$ absorption due to 
the hybridization mixing of the O $2p$ states with the Ce $4f$ states. 
This interpretation agrees with that for CeO$_2$ \cite{Allen85}.
Note that the low BE states, those at $1-2$ eV below $\rm E_F$, show 
another weak peak around $\sim 121$ eV, which is ascribed to the 
resonance of the localized Ce $4f$ electrons due to the Ce $4d 
\rightarrow 4f$ absorption. The very weak resonance of the low BE 
states indicates that the localized $4f$ states are almost unoccupied 
in LCeMO. On the other hand, the strong resonance of the O $2p$ 
electrons suggests that the valence-band states contain the extended 
states of $f$ symmetry, which are responsible for the non-negligible 
$4f$ population of $n_f$ in its ground state
($n_f \sim 0.5$ for CeO$_2$).
It is difficult to tell whether the resonance peak at $h\nu\sim 121$ 
eV for BEs $\approx 1-2$ eV has a Fano line-shape.
This is because the enhancement is very weak and it overlaps with
the strong La $5d$ resonance due to the La 4d $\rightarrow$ 4f RPES, 
which has a broad maximum around $h\nu=118-122$ eV \cite{Kang01}.

\begin{figure}[t]
\includegraphics[scale = 0.6]{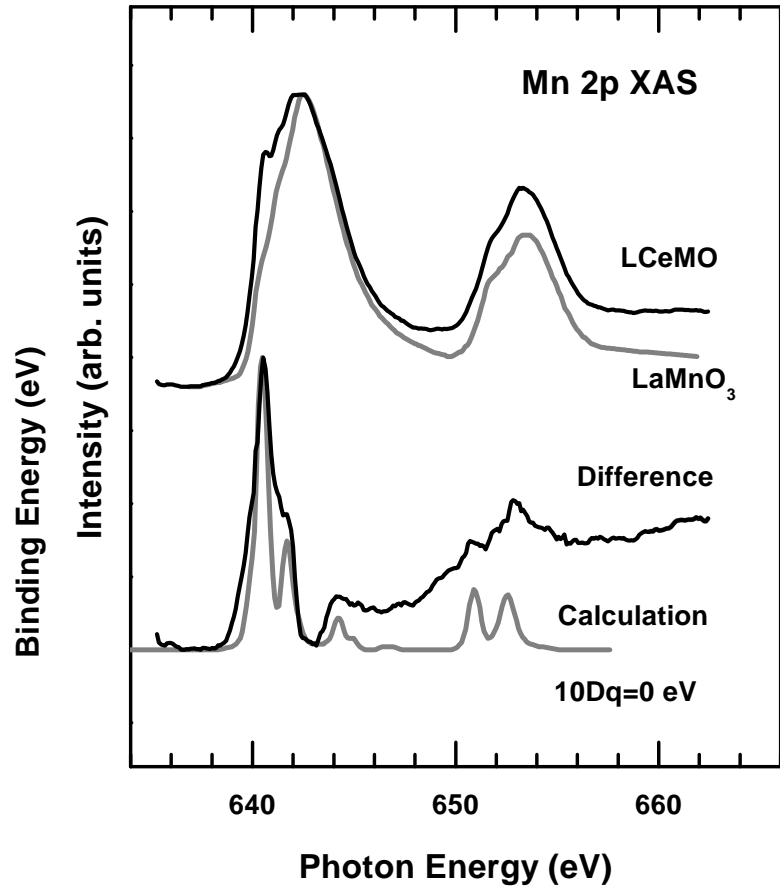}
\caption{Comparison of the Mn $2p$ XAS spectra of LCeMO and LaMnO$_3$
	(from Ref.\protect\cite{Mitra03b}).
	The difference curve between the Mn $2p$ XAS spectra of LCeMO 
	and LaMnO$_3$ is shown at the bottom as black lines. 
        The calculated Mn $2p$ XAS spectrum (gray lines) for 
	the Mn$^{2+}$ ion under the $O_h$ symmetry with $10Dq=0$ eV 
        (from Ref.\protect\cite{Groot90})	
	is compared to the difference curve. 
	See the text for the details.} 
\label{2pxas}
\end{figure}

In order to examine the resonating features more clearly,
we have compared the valence-band spectra around the two resonating
$h\nu$ values of $h\nu\approx 121$ eV and $h\nu\approx 124$ eV.
The left panel of Fig.~\ref{diff} compares the normalized valence-band 
spectra of LCeMO obtained for $h\nu=118$ eV and $h\nu=121$ eV,
and the difference between these two spectra is shown at the bottom.
Similarly, the right panel compares those for $h\nu=120$ eV and 
$h\nu=124$ eV, and the difference between these two spectra.
The left panel shows that the Ce $4f$ contribution is the largest 
at $\sim 1.5$ eV below E$_F$ with no Ce $4f$ emission at $\rm E_F$,
consistent with the insulating state above T$_C$. 
As mentioned in Fig.~\ref{rpes}, the magnitude of the Ce $4f$ resonance 
at $h\nu \approx 121$ eV is very weak, indicating that the localized 
Ce $4f$ sates are nearly unoccupied in LCeMO and that the Ce valence 
is far from $3+$. The right panel shows that the O $2p$ states resonate
at $h\nu\sim 124 $eV and are spread between $\sim 3-6$ eV BE with no 
states between $\rm E_F$ and 2 eV BE.  

Figure~\ref{2pxas} compares the Mn $2p$ XAS spectra of LCeMO 
and LaMnO$_3$ which was reproduced from Ref.~\cite{Mitra03b}.  
LaMnO$_3$ was chosen as the reference material which has the formally 
trivalent ($3+$) Mn ions and the same crystal symmetry.
The difference curve between the Mn $2p$ XAS spectrum of LCeMO 
and that of LaMnO$_3$ is shown at the bottom.  
The transition metal (T) $2p$ XAS spectrum results from the 
dipole transitions from the $2p$ core level to the empty $3d$ states.
The peak positions and the line shape of the T $2p$ XAS spectrum 
depend on the local electronic structure of the T ion, which provides 
the information about the valence state and the ground state symmetry 
of the T ion \cite{Groot90,Laan92}.
The Mn $2p_{3/2}$ and $2p_{1/2}$ spectral parts are clearly separated 
by the large $2p$ core-hole spin-orbit interaction.

\begin{figure}[t]
\includegraphics[scale = 0.6]{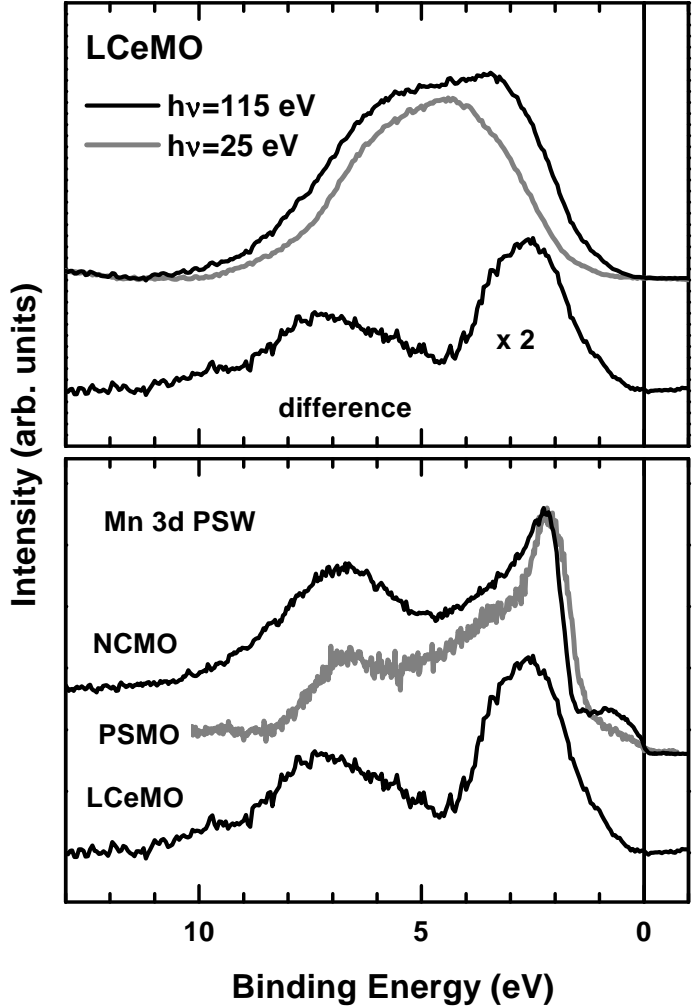}
\caption{Top: The extraction procedure for the Mn 3d PSW of LCeMO.
	See the text for the details.	
	Bottom: Comparison of the Mn $3d$ PSW of LCeMO to those of
	$7\%$-doped Nd$_{1/2}$Ca$_{1/2}$MnO$_3$ (NCMO)
	(from Ref.~\protect\cite{Kang03})
	and Pr$_{2/3}$Sr$_{1/3}$MnO$_3$ (PSMO) 
	(from Ref.~\protect\cite{Kang99}. 
	The former PSW for NCMO was obtained from the Mn $2p 
	\rightarrow 3d$ RPES.
	The extraction procedure for the latter PSW for PSMO is 
	described in Ref.~\protect\cite{Kang99}.}
\label{mn3d}
\end{figure}

\begin{figure}[t]
\includegraphics[scale = 0.6]{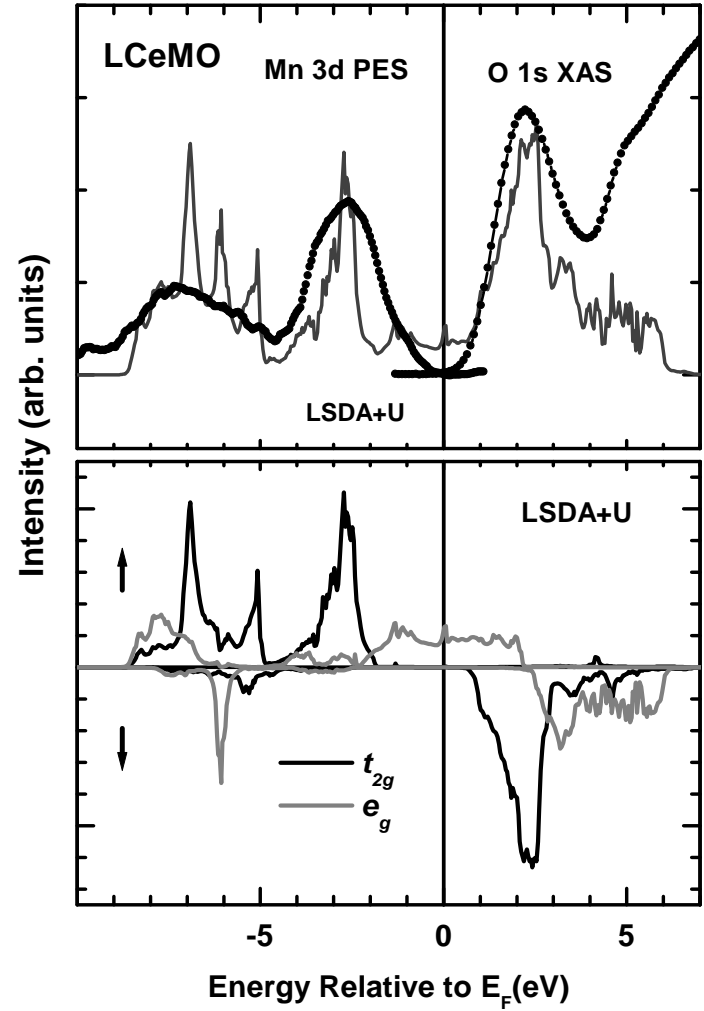}
\caption{Top: Comparison of the experimental Mn $3d$ PSW and O $1s$ XAS
	spectrum and the calculated Mn $3d$ PDOS obtained from 
	the LSDA$+U$ calculation.
	The O $1s$ XAS spectrum is shifted by referring to 
	the calculated PDOS. 
	Bottom: The calculated Mn $3d$ PDOSs of LCeMO obtained from
	the LSDA$+U$ calculation. The Mn $3d$ band is decomposed
	into $t_{2g}$ (black lines) and $e_g$ (gray lines) bands.}
\label{comp}
\end{figure}

Similarly as in other manganites \cite{Park96}, the Mn $2p$ XAS 
spectra of both LCeMO and LaMnO$_3$ exhibit two broad multiplets 
separated by the spin-orbit splitting of the Mn $2p$ core hole 
before subtraction. The main contributions to the broadening come from 
a large spread of the multiplets \cite{Saitoh95} and a covalent 
character of the ground state.
In contrast, the difference curve reveals the sharp structure 
around 640 eV, which resembles that in MnO \cite{Mitra03}, 
indicating the existence of a Mn$^{2+}$ component in LCeMO.
To check this argument, we have compared the difference curve to
the calculated Mn $2p$ XAS spectrum (gray lines) for the Mn$^{2+}$ ion 
under the octahedron symmetry ($O_h$) symmetry with $10Dq=0$ eV.
This result has been reproduced from Ref.~\cite{Groot90}.
This comparison clearly shows that the main features of the difference 
curve can be described mainly with the Mn$^{2+}$ component
under the $O_h$ symmetry. Thus Fig.~\ref{2pxas} confirms 
the Mn$^{2+}$-Mn$^{3+}$ mixed-valent state of Mn ions in LCeMO, 
in agreement with the previous finding \cite{Mitra03}. 

The upper panel of Fig.~\ref{mn3d} shows the extraction procedure 
of the Mn 3d partial spectral weight (PSW) of LCeMO. As a first 
approximation, the Mn $3d$ PSW of LCeMO is determined by subtracting 
the valence-band spectrum at h$\nu=25$ eV, where the O $2p$ emission 
is dominant, from that at the off-resonance (h$\nu=115$ eV) in Ce 
$4d \rightarrow 4f$ RPES, where the Mn $3d$ and O $2p$ emissions 
are comparable each other \cite{Kang01}.
The former spectrum has been scaled by a factor of $\sim 0.9$
to account for the h$\nu$-dependence of the O $2p$ photoionization
cross-section \cite{Yeh85}. The extracted Mn $3d$ states show two broad
structures, around $\sim 2$ eV and and $\sim 7$ eV. The spectral 
intensity at $\sim 1$ eV below $\rm E_F$ is weak.

The bottom panel of Fig.~\ref{mn3d} compares the Mn $3d$ PSWs
of LCeMO to those of Nd$_{0.5}$Ca$_{0.5}$Mn$_{0.93}$Cr$_{0.07}$O$_3$
(NCMO; Ref.~\cite{Kang03}) and 
Pr$_{2/3}$Sr$_{1/3}$MnO$_3$ (PSMO; Ref.~\cite{Kang99},
both of which are metallic in the ground state.
The former Mn $3d$ PSW for NCMO was obtained from the Mn $2p 
\rightarrow 3d$ RPES. 
The extraction procedure for the latter PSW for PSMO is described 
in Ref.~\protect\cite{Kang99}.
It was extracted by employing the same procedure as for LCeMO, 
where $h\nu=119$ eV was used as the off-resonance in Pr $4d \rightarrow
4f$ RPES and the $h\nu=18$ eV spectrum was considered as roughly 
representing the O $2p$ spectrum.
Indeed, the Mn $3d$ PSW for PSMO is very similar to that for NCMO, 
obtained from the Mn $2p \rightarrow 3d$ RPES, except for the 
slightly lower intensity near $\rm E_F$.   
This finding supports the validity of the extraction scheme
for the Mn $3d$ PSW employed in this work. 
 
In order to understand the microscopic origin of the valence-band 
electronic structures of LCeMO, we have compared the  
experimentally determined Mn $3d$ PSW and O $1s$ XAS spectrum
to the calculated Mn $3d$ PDOS, which was obtained from the supercell
LSDA$+U$ calculation.
The O $1s$ XAS spectrum reflects the transition from the O $1s$
core level to the unoccupied O $2p$ states hybridized to the other 
electronic states. Therefore the O $1s$ XAS provides a reasonable 
approximation for representing the unoccupied conduction-band 
electronic structure. The results are presented in Fig.~\ref{comp}.
The parameters used in this calculation are the Coulomb correlation 
$U=4.0$ eV and the exchange correlation $J=0.87$ eV for Mn $3d$
electrons, and $U=5.0$ eV and $J=0.95$ eV for Ce $4f$ electrons.
It is found that the LSDA$+U$ results are not very sensitive
to the $U$-value, within $\Delta U \approx \pm 1$ eV.
As shown at the bottom of Fig.~\ref{comp}, the LSDA$+U$ calculation
predicts a half-metallic ground state for LCeMO, which is consistent 
with the calculated electronic structure obtained in the virtual 
crystal approximation (VCA) \cite{Min01}. In this comparison, 
the O $1s$ XAS spectrum was shifted by $-527.5$ eV with reference 
to the to the LSDA$+U$ calculation \cite{shift}. 
This comparison shows that the measured PES/XAS data for LCeMO agrees 
reasonably well with the calculated Mn $3d$ PDOS, particularly 
in the peak positions. The broad peak around $-3$ eV originates from 
the t$_{2g}^3$ and e$_g$$^x$ majority-spin states, and the high BE
features ($-5 \sim -10$ eV) have the strongly mixed electron character 
of the Mn $3d$-O $2p$ electrons. 

The ground state of LCeMO becomes half-metallic in the LSDA$+U$ 
calculation, reflecting that the DE mechanism is still operative 
in LCeMO\cite{Lsda}. Indeed this prediction is consistent with 
the recent observation \cite{Mitra03b} of a tunneling magnetoresistance
in ferromagnetic tunnel junction built from 
LCeMO/La$_{0.7}$Ca$_{0.3}$MnO$_{3}$, which suggests a high degree 
of spin polarization in LCeMO. The LSDA$+U$ calculation, however, 
indicates that LCeMO would be a majority-spin carrier half-metal, 
contrary to a minority-spin carrier half-metal suggested by 
Ref. \cite{Mitra03b}. This point remains to be clarified experimentally
by employing direct spin-resolved experimental probes.
Finally, the measured spectral weight at $\rm E_F$ (I($\rm E_F$)) 
with respect to the Mn $t_{2g}$ peak is lower than the calculated PDOS 
at $\rm E_F$ (N($\rm E_F$)). This difference near $\rm E_F$ can be 
ascribed partly due to the measurement temperature being above the M-I 
transition and also due to some kind of the carrier localization 
mechanism. The most plausible localization mechanism would be the small 
polaron formation induced by the strong $e_g$ electron-phonon 
interaction, originating from the Jahn-Teller active Mn$^{3+}$ ion 
in LCeMO \cite{Kang99,Lee97}.

\section{Conclusion}

We have performed PES and XAS measurements for LCeMO thin film,
including the RPES measurement near the Ce $4d \rightarrow 4f$ 
absorption edge, and the XAS measurements near the Ce and La $3d$, 
Mn $2p$, and O $1s$ absorption edges. The PES and XAS data have been 
compared to the band-structure calculations performed in the LSDA$+U$ 
method. Both the Ce $3d$ core-level PES and XAS spectra are very 
similar to those of CeO$_2$ with Ce$^{4+}$, indicating that 
the Ce valence is far from $3+$ in LCeMO.
The RPES measurement also shows very weak enhancement around
the Ce $4d \rightarrow 4f$ absorption edge, providing evidence
that the localized Ce $4f$ states are almost empty in the ground state.
The Mn $2p$ XAS spectrum reveals the multiplet features of Mn$^{2+}$
ions besides those of Mn$^{3+}$ ions, confirming the existence of 
the Mn$^{2+}$-Mn$^{3+}$ mixed-valent states of Mn ions in LCeMO.
The comparison between the measured Mn $3d$ PES/XAS data and the 
calculated Mn $3d$ PDOS in the LSDA$+U$ method shows a reasonably good 
agreement, particularly in the peak positions.
The LSDA$+U$ calculation predicts a half-metallic ground state 
for LCeMO with majority spin carriers at $\rm E_F$.

\begin{acknowledgements}

This work was supported by the KRF (KRF--2002-070-C00038) and
by the KOSEF through the CSCMR at SNU and the eSSC at POSTECH.
The PAL is supported by the MOST and POSCO in Korea.

\end{acknowledgements}


\begin{thebibliography}{}
\bibitem{int} The intensity of each peak is proportional to square 
	of the overlap integral between the initial and final states.
\bibitem{shift} There is some uncertainty in choosing the zero 
	position from the XAS spectrum.
\bibitem{Lsda} The supercell calculation in the LSDA yields the
	normal metallic LCeMO, contrary to the LSDA$+U$ result. 
	This implies that including the $U$-effect is essential to 
	describe the half-metallic nature of LCeMO.
\end{thebibliography}
\end{document}